\documentclass[fleqn,usenatbib]{mnras}

\usepackage{newtxtext,newtxmath}
\usepackage[T1]{fontenc}
\usepackage{ae,aecompl}

\usepackage{graphicx}	% Including figure files
\usepackage{amsmath}	% Advanced maths commands
\usepackage{cleveref}
\usepackage{multirow}
\usepackage{placeins}
\usepackage{subcaption} % Enables subfigures

\DeclareMathOperator{\Tr}{Tr}

\title[RFI Mitigation via Likelihood Reweighting]{Enhanced Bayesian RFI Mitigation and Transient Flagging Using Likelihood Reweighting}

\author[D.Anstey et al.]{
Dominic Anstey,$^{1,2}$\thanks{E-mail: da401@cam.ac.uk}
Samuel A. K. Leeney,$^{1,2}$
\\
$^{1}$Astrophysics Group, Cavendish Laboratory, J. J. Thomson Avenue, Cambridge, CB3 0HE, UK\\
$^{2}$Kavli Institute for Cosmology, Madingley Road, Cambridge, CB3 0HA, UK
}

\date{Accepted XXX. Received YYY; in original form ZZZ}

\pubyear{2023}

\begin{document}
\label{firstpage}
\pagerange{\pageref{firstpage}--\pageref{lastpage}}
\maketitle
\begin{abstract}
Contamination by Radio Frequency Interference (RFI) is a ubiquitous challenge for radio astronomy. In particular, transient RFI is difficult to detect and avoid, especially in large data sets with many time bins. In this work, we present a Bayesian methodology for time-dependent, transient anomaly mitigation. In general, the computation time for correcting for transient anomalies in time-separated data sets grows proportionally with the number of time bins. We demonstrate that utilising likelihood reweighting can allow our Bayesian anomaly mitigation method to be performed with a computation time close to independent of the number of time bins. In particular, we identify a factor of 25 improvement in computation time for a test case with 2000 time bins. We also demonstrate how this method enables the flagging threshold to be fit for as a free parameter, fully automating the mitigation process. We find that this threshold fitting also prevents overcorrecting of the data in the case of wide priors. Finally, we investigate the potential of the methodology as a transient detector. We demonstrate that the method is able to reliably flag an individual anomalous data point out of 302,000 provided the SNR $>10$. 
\end{abstract}


%%%%%%%%%%%%%%%%%%%%%%%%%%%%%%%%%%%%%%%%%%%%%%%%%%

%%%%%%%%%%%%%%%%% BODY OF PAPER %%%%%%%%%%%%%%%%%%

\section{Introduction}\label{sec:intro}
The field of radio astronomy has been rapidly growing in terms of both reach and data complexity. The number of known radio sources has been exponentially increasing, and will increase further in the future with the development of the SKA \citep{bourke2015advancing} and the ngVLA \citep{mckinnon2019ngvla}. As the volume of information on the sky increases, Astronomers also seek increasingly faint signals requiring more sensitive instruments and advanced data analysis techniques. 

With the development of modern telecommunications devices, data from radio telescopes is becoming increasingly contaminated with interfering signals such as Radio Frequency Interference (RFI), which has become very difficult to avoid entirely\citep{fridman2001rfi}, except in observations from extremely remote locations \citep{monsalve2023}. This problem is worsened by the fact that current signals of interest, such as the Global 21cm signal \citep{bowman18, de2022reach, monsalve2023, singh2022}, lie in the same unprotected frequency bands as said devices.

RFI can emanate from a range of human-made sources, such as communication devices, satellites, radar systems, etc. It poses a significant challenge to radio astronomy, as it can obscure or mimic genuine celestial signals. RFI can be constant in time or transient \citep{czech2018cnn}. Transient RFI is particularly problematic because it is hard to detect and thus much more difficult to avoid. The SKA will gather up to one 1TB of data per second \citep{scaife2020big}. With the volume of data to be analysed so large and the level of complex contaminants so high, there is a serious need for new data analysis techniques that are highly efficient and sensitive to such transient events. 

The are various proposed ways to mitigate RFI. \cite{2010MNRAS.405..155O} use post-correlation classification methods in \textsc{AOFLAGGER}, which is used by the LOFAR \citep{rottgering2003lofar}. The FAST \citep{nan2006five} telescope uses spatial filtering techniques \citep{wang2022description}. More recently, deep learning methods have been utilised \citep{kerrigan2019optimizing}. For a more in depth review of the current literature, we recommend \citep{ford2014rfi} or \citep{baan2019implementing}. There are few methods designed specifically for transient RFI detection, as noted by \citep{czech2018dictionary}, who proposes a dictionary based approach to transient RFI detection.

Transient RFI is exceptionally problematic when the signal of interest is itself transient. For example a transient RFI burst could not only obscure a signal (such as a Fast Radio Burst (FRB)) but could also mimic it leading to a false detection \citep{cendes2018rfi}. This problem can be partially addressed using spectral kurtosis \citep{nita2019statistical}. However, spectral kurtosis is inadequate in various cases as described in \citet{smith2022simulating}. Furthermore, with many modern projects utilizing Bayesian methods in their data analysis pipelines, there is a urgent need for Bayesian RFI mitigation techniques.

In this paper, we present a novel Bayesian methodology for efficient transient RFI mitigation. In \Cref{sec:methods} we define the method. In \Cref{sec:LST_flagging_results} we test our methods on a simple toy model. In \Cref{sec:transient_detection} we evaluate these methods when used to be used to locate transient signals themselves, as well as mitigate against them. In \Cref{sec:conclusions}, we present our conclusions.

\section{Methods}\label{sec:methods}
\subsection{Bayesian Anomaly Mitigation}\label{sec:flagging}
\citet{leeney22} proposed a fully Bayesian methodology for simultaneous anomaly flagging and excision. The term \textit{likelihood}, in the context of a single data point, defines the probability of observing that data point $\mathcal{D}_i$ given some model and its constituent parameters $\mathcal{M}_i\left(\theta\right)$. This assumes a predefined probability distribution for the noise. For instance, assuming Gaussian noise results in a likelihood of the form 
\begin{equation}\label{eq:single_L}
     \log \mathcal{L}_i\left(\theta\right) = -\frac{1}{2}\log\left(2\pi\sigma_\mathrm{n}^{2}\right) - \frac{1}{2}\left(\frac{\mathcal{D}_i - \mathcal{M}_i\left(\theta\right) }{\sigma_\mathrm{n}}\right)^{2}
\end{equation}
where $\theta$ is the parameters, and $\sigma_\mathrm{n}$ is the noise amplitude. Alternatively, for a data set containing $N_i$ points and a corresponding model, with a single parameter set $\theta$ for all points, assuming no correlations, the overall likelihood is the product of the probabilities of each datum, giving
\begin{equation}\label{eq:multi_L}
    \log \mathcal{L}\left(\theta\right) = \sum_i^{N_i} \log \mathcal{L}_i\left(\theta\right).
\end{equation}
However, this likelihood cannot account for corrupted data points, such as those contaminated by RFI. Typically, anomalies are flagged and excised prior to the Bayesian fitting procedure. This can be problematic, as it leads to potentially useful information being thrown away. \citet{leeney22} showed that anomalies can be modeled simultaneously in a Bayesian fashion. They do this using a piece-wise likelihood capable of modelling both the probability of abnormality and the probability that each datum fits to the model of interest.

\citet{leeney22} then defined a prior probability on corruption, $p$, and the two possibilities in the piece-wise likelihood are compared for each datum. Data considered less likely to fit the underlying model than some threshold (defined by the prior on abnormality) are considered more likely to be anomalous. It is assumed that the anomalous points are uncorrelated with the underlying data structure, instead falling in an amplitude range $[0-\Delta]$.

 A Bayesian likelihood in which the model is unaffected by anomalous points can be created by predicting how likely each datum is to be contaminated. If at any point data is predicted to be anomalous, an Occam's penalty is applied. This is because the Bayesian evidence favours the simplest solution that best describes the data, so without the penalty the `most likely' solution would be to flag all of the data as anomalous.

 Ideally, the likelihood should then be marginalised over all possible arrangements of data point flags. However in practice, the optimal arrangement can be approximated. This is achieved by comparing, for each data point, the value of the penalised full likelihood compared to the fixed penalty and taking the larger. The resulting likelihood then has the form
\begin{multline}\label{eq:time_av_flagged}
\log \mathcal{L} \left(\theta\right) = \sum_i \left[\log \mathcal{L}_i \left(\theta\right) + \log\left(1-p_i\right)\right]\epsilon_i + 
\\\left[ \log p_i - \log \Delta \right]\left(1-\epsilon_i\right)
\end{multline}
where $\epsilon_i$ is a mask of anomalous points defined as 
\begin{equation}\label{eq:RFI_mask}
    \epsilon_i = 
    \begin{cases}
        1, &
    \begin{aligned}
        &[\log{\mathcal{L}_i} + \log({1-p_i}) \\
        &> \log p_i - \log \Delta]
    \end{aligned}\\ 
        0, & \text{otherwise}.
    \end{cases}
\end{equation}

\citet{leeney22} showed that this likelihood enabled anomalous data points to be efficiently flagged and corrected for automatically in a Bayesian model fit of a 1D data set. The focus of their work is extending this methodology for application to 2D, time-varying data sets.

\subsection{Time Binned Modelling}\label{sec:LST_fitting}
\citet{anstey22} proposed a methodology for efficient inclusion of time dependent data sets in Bayesian modelling. Typically, in the case of time-varying data, modelling each time bin separately is unfeasible as it requires a complete model with its own set of parameters for each bin, which can quickly result in the dimensionality of the fit becoming very large for more than a few time bins. As a result, time-varying data is typically modelled by fitting a model to the time averaged data set. For example, in the case of the Gaussian likelihood described in \Cref{eq:single_L,eq:multi_L}, a time-averaged likelihood would take the form

\begin{equation}\label{eq:time_av_full_L}
    \log \mathcal{L} \left(\theta\right) = \sum_i -\frac{1}{2}\log\left(2\pi\sigma_\mathrm{n}^{2}\right) - \frac{1}{2}\left(\frac{\frac{1}{N_t}\sum_j \mathcal{D}_{ij} - \mathcal{M}_i\left(\theta\right) }{\sigma_\mathrm{n}}\right)^{2},
\end{equation}
where $j$ indexes time bins and $N_t$ is the total number of time bins.

However, in \citet{anstey22}, the case was considered where a time varying model can be defined as a product of a parameter-dependent component and a parameter-independent component, where only the parameter-independent component has time dependence, of the form
\begin{equation}\label{eq:model_breakdown}
    \mathcal{M}_{ij} \left(\theta\right) = \mathcal{F}_i \left(\theta\right) \mathcal{G}_{ij}.
\end{equation}

If it is the case that the time-dependant component $G_{ij}$, despite being independent of the specific value of the parameters, is required to be different for the different parameters in $\theta$, which will be the case if different parameters are describing different effects, this can be expressed more fully as 

\begin{equation}\label{eq:model_breakdown_w_k}
    \mathcal{M}_{ij} \left(\theta_k\right) = \mathcal{F}_i \left(\theta_k\right) \mathcal{G}_{ijk},
\end{equation}
where $k$ indexes over the parameters and $\theta_k$ expresses the individual elements of the vector $\theta$.

An example of a case where a model of this form can be defined is for a set of parameters defining an astrophysical observable that is constant on the time scales of the experiment, with time variance only being introduced by the rotation of the Earth. Another possibility would be a data set of just noise with transient signals, which would simply have a model of $\mathcal{M}_{ij}\left(\theta\right) = 0$, which still satisfies this condition.

If such a model can be defined, it becomes possible to implement simultaneous fitting of separate time bins of data to corresponding models without the dimensionality of the parameters increasing, as each time bin fits for the same parameter set. Such a fitting process can be implemented by modifying the likelihood shown in \Cref{eq:time_av_full_L} to
\begin{equation}\label{eq:time_sep_full_L}
    \log \mathcal{L} \left(\theta\right) = \sum_{ij} -\frac{1}{2}\log\left(2\pi\sigma_\mathrm{n}^{2}\right) - \frac{1}{2}\left(\frac{\mathcal{D}_{ij} - \mathcal{M}_{ij}\left(\theta\right) }{\sigma_\mathrm{n}}\right)^{2}.
\end{equation}

It was demonstrated in \citet{anstey22}, in the context of Global 21cm experiments, that using this full time-dependent likelihood in a Bayesian model fit enabled the time variance of the model to be exploited to constrain the model parameter more tightly than could be achieved for a time-averaged model.

\subsection{Time Binned Anomaly Mitigation}\label{sec:LST_flagging}
Applying the Bayesian anomaly mitigation technique described in \Cref{sec:flagging} to the time dependent likelihood described in \Cref{sec:LST_fitting} is straightforward, requiring only that the flagged likelihood shown in \Cref{eq:time_av_flagged} be extended into 2 dimensions as

\begin{multline}\label{eq:time_sep_flagged}
\log \mathcal{L} \left(\theta\right) = \sum_{ij} \left[\log \mathcal{L}_{ij} \left(\theta\right) + \log\left(1-p_{ij}\right)\right]\epsilon_{ij} + 
\\
\left[ \log p_{ij} - \log \Delta \right]\left(1-\epsilon_{ij}\right)
\end{multline}
where $\epsilon_{ij}$ is a mask of anomalous points defined as 
\begin{equation}\label{eq:RFI_mask_sep}
    \epsilon_{ij} = 
    \begin{cases}
        1, &
    \begin{aligned}
        &[\log{\mathcal{L}_{ij}} + \log({1-p_{ij}}) \\
        &> \log p_{ij} - \log \Delta]
    \end{aligned}\\ 
        0, & \text{otherwise},
    \end{cases}
\end{equation}
and $\log{\mathcal{L}_{ij}}$ is the likelihood of a single data point in a single time bin
\begin{equation}\label{eq:time_sep_single_L}
    \log \mathcal{L}_{ij} \left(\theta\right) = -\frac{1}{2}\log\left(2\pi\sigma_\mathrm{n}^{2}\right) - \frac{1}{2}\left(\frac{\mathcal{D}_{ij} - \mathcal{M}_{ij}\left(\theta\right) }{\sigma_\mathrm{n}}\right)^{2}.
\end{equation}

However, implementing this method in practice faces a challenge. The likelihood shown in \Cref{eq:time_sep_full_L} requires a summation over time bins. This means the computation time of the likelihood grows linearly with the number of time bins used in the data set. As a result, larger numbers of time bins can greatly slow the fitting procedure. 

In \citet{anstey22}, this issue was resolved by reformatting the likelihood such that all summations over time could be calculated once, outside of the likelihood, removing the dependence of the calculation time on the number of time bins. However, this solution is not possible to implement when the anomaly correcting procedure is also implemented. This is because the value of $\log{\mathcal{L}_{ij}}$ must be calculated for every time bin within the likelihood, in order to evaluate $\epsilon_{ij}$, as shown in \Cref{eq:RFI_mask_sep}. Therefore, an alternative method of speeding the likelihood evaluation is needed to make this process viable in practice. This can be achieved using likelihood reweighting.

\subsection{Likelihood Reweighting}\label{sec:LRW}
The process of likelihood reweighting was pioneered in the context of gravitational waves \citep{payne19, romeroshaw19}. It is a method for speeding the evaluation of a posterior and evidence in a Bayesian fit for the case of a complex model that is otherwise slow to evaluate. This process relies on several key criteria.

Firstly, two models, $\mathcal{M}_\mathrm{F}\left(\theta\right)$ and $\mathcal{M}_\mathrm{S}\left(\theta\right)$, are required. One of these must be quick to evaluate, which will henceforth be assumed to be model F, and one is slower to evaluate, which we define as model S. These two models must be parameterised by the same parameter vector $\theta$, with the same prior distribution $\mathcal{\pi}\left(\theta\right)$. They must also have their posterior peak in approximately the same region of parameter space. 

By definition, the posteriors of the two models can be expressed as 
\begin{equation}\label{eq:post_F}
    \mathcal{P}_\mathrm{F} \left(\theta | \mathcal{D}, \mathcal{M}_\mathrm{F}\right) = \frac{\mathcal{L}_\mathrm{F}\left(\mathcal{D} | \theta, \mathcal{M}_\mathrm{F}\right)\mathcal{\pi}\left(\theta\right)}{\mathcal{Z}_\mathrm{F}}
\end{equation}
and
\begin{equation}\label{eq:post_S}
    \mathcal{P}_\mathrm{S} \left(\theta | \mathcal{D}, \mathcal{M}_\mathrm{S}\right) = \frac{\mathcal{L}_\mathrm{S}\left(\mathcal{D} | \theta, \mathcal{M}_\mathrm{S}\right)\mathcal{\pi}\left(\theta\right)}{\mathcal{Z}_\mathrm{S}},
\end{equation}
where $\mathcal{L}_\mathrm{F}\left(\mathcal{D} | \theta, \mathcal{M}_\mathrm{F}\right)$ and $\mathcal{L}_\mathrm{S}\left(\mathcal{D} | \theta, \mathcal{M}_\mathrm{S}\right)$ are the likelihoods calculated from the two models and $\mathcal{Z}_\mathrm{F}$ and $\mathcal{Z}_\mathrm{S}$ are the respective evidences. 

However, given the aforementioned criterion that the two models have the same priors, the prior can be expressed in terms of model F, as
\begin{equation}\label{eq:prior_as_F}
    \mathcal{\pi}\left(\theta\right) = \frac{\mathcal{Z}_\mathrm{F} \mathcal{P}_\mathrm{F}\left(\theta | \mathcal{D}, \mathcal{M}_\mathrm{F}\right)}{\mathcal{L}_\mathrm{F}\left(\mathcal{D} | \theta, \mathcal{M}_\mathrm{F}\right)}
\end{equation}
and substituted into \Cref{eq:post_S} to give
\begin{equation}\label{eq:LRW_full}
    \mathcal{P}_\mathrm{S} \left(\theta | \mathcal{D}, \mathcal{M}_\mathrm{S}\right) = \mathcal{P}_\mathrm{F}\left(\theta | \mathcal{D}, \mathcal{M}_\mathrm{F}\right) \frac{\mathcal{L}_\mathrm{S}\left(\mathcal{D} | \theta, \mathcal{M}_\mathrm{S}\right)}{\mathcal{L}_\mathrm{F}\left(\mathcal{D} | \theta, \mathcal{M}_\mathrm{F}\right)} \frac{\mathcal{Z}_\mathrm{F}}{\mathcal{Z}_\mathrm{S}}.
\end{equation}
Thus the posterior of model S can be evaluated without having to perform a full Bayesian model fit with the slow-to-calculate likelihood. This is achieved by instead performing a model fit of the much faster evaluated model F. Given the criterion that the bulk of the two models posteriors occupy similar regions of the parameter space, the samples of this fast evaluated posterior will cover the same parameter volume as a hypothetical posterior of the slower model. Therefore, reweighting the posterior samples of model F by a factor of the ratio of the likelihoods
\begin{equation}\label{eq:reweights}
    w\left(\theta\right) = \frac{\mathcal{L}_\mathrm{S}\left(\mathcal{D} | \theta, \mathcal{M}_\mathrm{S}\right)}{\mathcal{L}_\mathrm{F}\left(\mathcal{D} | \theta, \mathcal{M}_\mathrm{F}\right)},
\end{equation}
will convert the samples to samples of the posterior of model S, to within a constant factor given by the ratio of the evidence. By this method, a set posterior for model S can be evaluated with the slow likelihood only needing to be computed for the relatively small number of posterior samples. This can be significantly faster than sampling the complex likelihood across the entire prior volume. For a particularly slow to calculate likelihood, this can make the fitting procedure significantly faster. This effect will be demonstrated in \Cref{sec:comp_time}.

Therefore, the process of likelihood reweighting is well suited to achieving fast and efficient time dependent anomaly mitigation. By taking the time averaged anomaly correcting likelihood as the quickly evaluated model F and the full time separated likelihood as the slowly evaluated model S, the full time dependent likelihood posterior can be evaluated quickly, allowing the Bayesian anomaly mitigation procedure to be implemented practically on time dependent data sets as follows. For the slow likelihood, the likelihood of each data point is calculated according to \Cref{eq:time_sep_flagged,eq:RFI_mask_sep,eq:time_sep_single_L}.

For the fast likelihood, however, in order to preserve as much of the time-dependent data as possible, the following process is implemented. First, outside the likelihood, the time average of the data set and model are evaluated as
\begin{equation}
    \widetilde{D_{i}} = \frac{1}{N_{t}} \sum_i D_{ij}
\end{equation}
and
\begin{equation}
    \widetilde{M_{i}}\left(\theta\right) = \mathcal{F}_i \left(\theta\right) \frac{1}{N_{t}} \sum_i G_{ij}
\end{equation}
respectively.

In addition, a further set of time-summed quantities are calculated, as defined in Section 2.1 of \cite{anstey22},

\begin{itemize}
    \item $T_{\mathrm{D}\, i} = \sum_j D_{ij}$
    \item $G_{ik} = \sum_j G_{ijk}$
    \item $T_{{\mathrm{D} \cdot \mathrm{G}}\, ik} = \sum_j D_{ij} G_{ijk}$
    \item $G_{\mathrm{sq}\, i\,k} = \sum_j {G_{ijk}}^2$
    \item $G_{\mathrm{cross}\, i\,k_1\, k_2} = \sum_j G_{k_1}\left(\nu_i, t_j\right) G_{k_2}\left(\nu_i, t_j\right)$
\end{itemize}

Then, the likelihood of each time-averaged data point can be calculated according to
\begin{equation}\label{eq:fast_single_L}
     \log \mathcal{L}_i\left(\theta\right) = -\frac{1}{2}\log\left(2\pi\sigma_\mathrm{n}^{2}\right) - \frac{1}{2}\left(\frac{\widetilde{D_{i}} - \widetilde{M_{i}}\left(\theta\right) }{\sigma_\mathrm{n}}\right)^{2}
\end{equation}
and a threshold array calculated as
\begin{equation}\label{eq:fast_RFI_mask}
    \widetilde{\epsilon_i} = 
    \begin{cases}
        1, &
    \begin{aligned}
        &[\log{\mathcal{L}_i} + \log({1-p_i}) \\
        &> \log p_i - \log \Delta]
    \end{aligned}\\ 
        0, & \text{otherwise}.
    \end{cases}
\end{equation}
The total number of flagged channels can then be calculated as
\begin{equation}\label{eq:fast_number_flagged}
    N_\mathrm{flagged} = \sum_i \widetilde{\epsilon_i}
\end{equation}
From there, the full, fast likelihood can be evaluated, according to Equation 7 of \citet{anstey22}, as
\begin{multline}\label{eq:fast_L}
        \log\mathcal{L}_\mathrm{F}\left(\theta\right) = -\frac{1}{2} N_\mathrm{flagged} N_{x} \log\left(2\pi \sigma_\mathrm{n} \right) \\ -\frac{\sum_i T_{\mathrm{D}\, i} \widetilde{\epsilon_i}}{2{\sigma_\mathrm{n}}^2}  - \frac{1}{{2\sigma_\mathrm{n}}^2} \left[\sum_i \sum_k G_{\mathrm{sq}\, i\,k} {F_i\left(\theta_k\right)}^2 \widetilde{\epsilon_i} + \right.\\ \left. \sum_{k_1} \sum_{k_2}\left[ \sum_i G_{\mathrm{cross}\, i\,k_1\, k_2} F_i\left(\theta_{k_1}\right) F_i\left(\theta_{k_2}\right)\widetilde{\epsilon_i} \right] \right.\\ \left. - \Tr{\sum_i G_{\mathrm{cross}\, i\,k_1\, k_2} F_i\left(\theta_{k_1}\right) F_i\left(\theta_{k_2}\right)\widetilde{\epsilon_i}} \right] \\
        + \frac{1}{{\sigma_\mathrm{n}}^2} \sum_i \sum_k T_{{\mathrm{D} \cdot \mathrm{G}}\, ik} F_i\left(\theta_k\right)\widetilde{\epsilon_i} \\
        + N_\mathrm{flagged} N_t \log\left(1-p_i\right) \\
         +\left(N_x - N_\mathrm{flagged}\right) N_t \left[\log p_i - \log\Delta\right].
\end{multline}

This likelihood uses the methodology described in \citet{anstey22} to evaluate a time-dependent likelihood in a time-independent fashion, whilst simultaneously implementing the anomaly mitigation methodology in a time-independent fashion by flagging out entire channels where contamination occurs, rather than only the contaminated time bins of those channels. 

In the following sections, this complete process will be tested on simulated data to evaluate its efficacy.

\section{Time Dependant RFI Mitigation}\label{sec:LST_flagging_results}
\subsection{Toy Model}\label{sec:toy_model}
In order to evaluate the efficacy of this method at correcting for the impact of transient RFI in time separated data, a toy simulated time dependant data set was generated of the form
\begin{equation}\label{eq:data_toy_model}
    \mathcal{D}_{ij} = \left[ \alpha_j \sin \left(\omega_jx_i +\phi_j\right )+ \gamma_j\right ]x_i^{-2.55} + \hat{\sigma},
\end{equation}
where $x$ describes the data channel (e.g frequency channel) value, indexed by i, and $\alpha_j$, $\omega_j$, $\phi_j$ and $\gamma_j$ are time-dependant variables that are chosen to vary gradually over the time bins. For each of the four values, a start point is randomly chosen uniformly from the range [0-40] for $\gamma_j$ and [0-1] for the other variables, and a step size is chosen uniformly in the range [0-5] for $\gamma_j$ and [-0.05 - 0.05] for all others. The 4 variables are then set to be 
\begin{equation}\label{eq:toy_model_vars}
    \mathrm{variable}_j = \mathrm{start} + j \times \mathrm{step}.
\end{equation}
$\hat{\sigma}$ is a realisation of random Gaussian white noise added to the data, which, unless otherwise specified, was set to have a standard deviation of 0.25. A different noise realisation is added to each time-bin.

A toy model of this form was chosen as it approximates the form of data from a global 21cm experiment, with the power law with a 2.55 spectral index approximating the spectral variation of the diffuse emission from the sky and the time varying sinusiods approximating the convolution of the diffuse emission with a chromatic antenna beam, as the Earth rotates. This is described, for example, in Equation 18 of \citet{anstey21} and seen in \citet{bowman18}. Global 21cm experiments are an anticipated use case of this process, so a toy model of this form enables these experiments to be used as a test case of this process.

In addition, a simulated data set of this form has the required structure to apply the time-separated model fitting as specified in \Cref{eq:model_breakdown}, with
\begin{equation}\label{eq:toy_model_G}
    \mathcal{G}_{ij} = \alpha_j \sin \left(\omega_jx_i +\phi_j\right )+ \gamma_j
\end{equation}
and
\begin{equation}\label{eq:toy_model_F}
      \mathcal{F}_i \left(\theta\right) = x_i^{-\theta}
\end{equation}
with a `true value' of $\theta=2.55$. 

\begin{figure}
    \centering
    \includegraphics[width=\columnwidth]{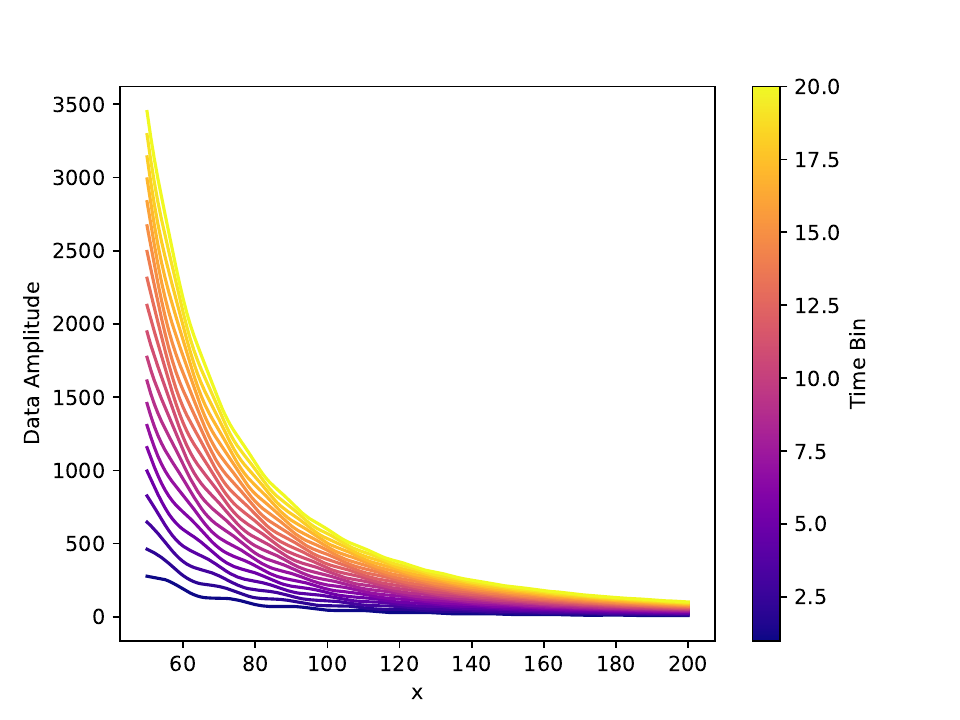}
    \caption{Example test data set with 20 time bins generated according to the toy model described in \Cref{sec:toy_model}.}
    \label{fig:example_data_set}
\end{figure}

\Cref{fig:example_data_set} shows an example simulated data set generated for 20 time bins using this toy model. Once such a data set has been generated, any arrangement of anomalous points can be then be added in order to test the proposed Bayesian anomaly mitigator.

\subsection{Parameterising the Threshold}\label{sec:threshold_param}
In \citet{leeney22}, the likelihood threshold value for determining if a point should be flagged as an anomaly or not, $p$, was set to a fixed value. However, doing so produces a challenge when applied to time-dependant modelling.

As shown in \Cref{eq:time_sep_flagged} and \Cref{eq:RFI_mask_sep}, the general method of operation for the Bayesian anomaly correction procedure is to calculate the likelihood of each separate data point, compare it to a predefined threshold value and if it exceeds the threshold, include it with an appropriate weighting and if it does not, add a fixed penalty to the total likelihood. This is acceptable under the assumption that only anomalous points will have likelihoods lower than the threshold. 

\begin{figure}
    \centering
    \includegraphics[width=\columnwidth]{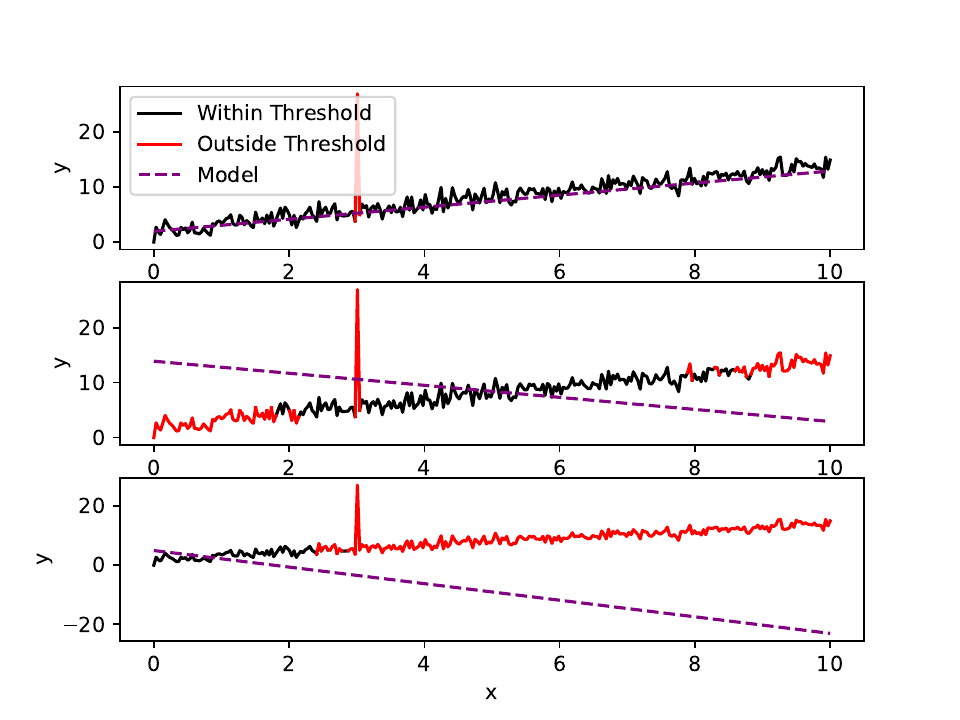}
    \caption{Example, for linear noisy data, of the range of data points that are further from the model than a defined threshold for a model that closely matches the data set (top), moderately differs from the data set (middle) and significantly differs from the data set (bottom).}
    \label{fig:thresh_example}
\end{figure}

However, in cases where the model used has relatively large priors and significant variability, it becomes possible for a given parameter sample to produce a model sufficiently different from the data set that significant numbers of data points have likelihoods below the threshold, as demonstrated by \Cref{fig:thresh_example}. 

This becomes of concern if the priors are wide enough compared to the posterior peak to result in sections of the prior space in which every data point has a likelihood below the defined threshold set by $p$, as, in this case, the overall likelihood is a sum of only fixed penalty terms, and thus is constant. As a result, all sections of the prior space that satisfy this condition have the same likelihood, which manifests as the outer boundaries of the likelihood surface becoming flat.

If the majority of the prior space is not flattened in this manner, this has minimal impact on the process of performing a model fit and recovering the parameter posteriors. However, if the likelihood surface is particularly steep, it is possible for the majority of the space to become flat, except for a very narrow region around the posterior peak. In this case, performing a model fit stops being possible, as there exists no variation to guide an algorithm towards the peak. 

\begin{figure}
    \centering
    \includegraphics[width=\columnwidth]{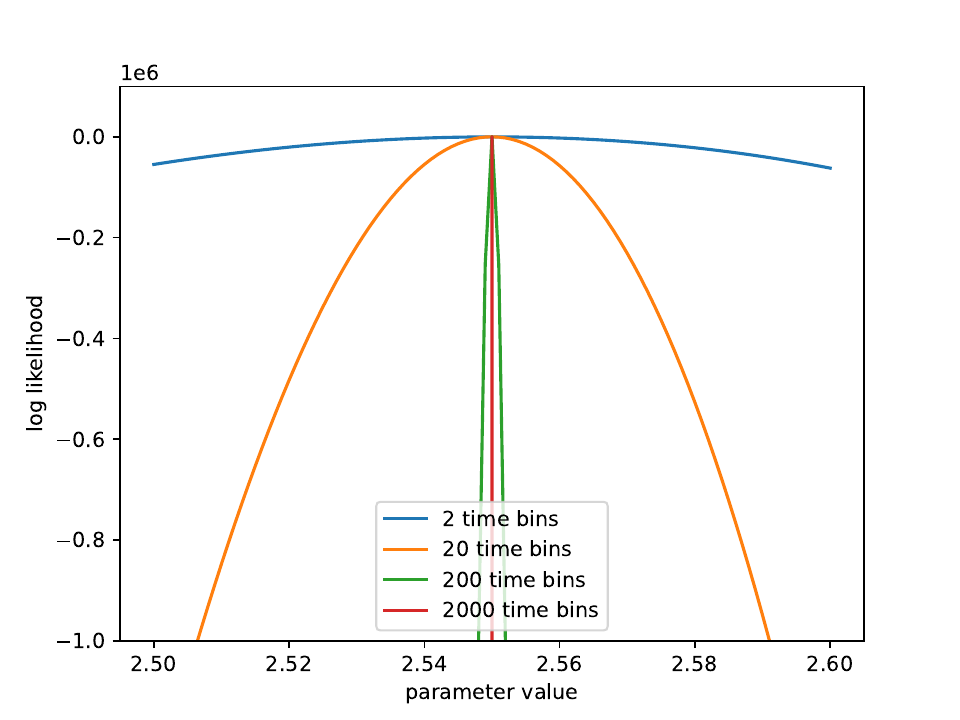}
    \caption{Plot of the total summed log likelihood shown in \Cref{eq:time_sep_full_L} for the toy model described in \Cref{sec:toy_model} as a function of parameter value for 2, 20, 200, and 2000 time bins.}
    \label{fig:time_dep_likelihood_curves}
\end{figure}

This is a challenge when performing time-dependant model fits, because the additional time information amplifies the effect of model disagreement. A parameter sample that gives a model that matches well to a data set with many time bins will give a high likelihood for each data point of each time bin, and thus give a larger overall likelihood than for an equally well matching data set with fewer time bins, and vice versa. As demonstrated in \Cref{fig:time_dep_likelihood_curves} for the proposed 1-parameter toy model, defined by \Cref{eq:toy_model_F,eq:toy_model_G}, with 2, 20, 200, and 2000 time bins, this has the effect of steepening the likelihood surface, the more time bins the data set has. 

\begin{figure*}
    \centering
    \begin{subfigure}[b]{0.55\textwidth}
         \centering
         \includegraphics[width=\textwidth]{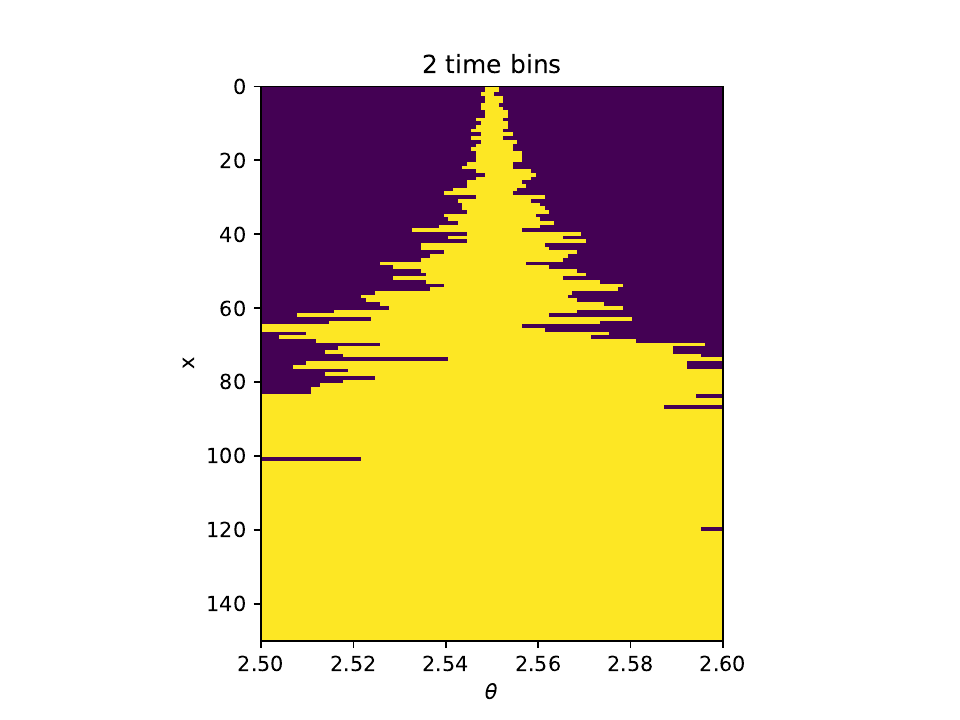}
         \label{subfig:2t_loglike_surface}
     \end{subfigure}
     \hspace{-7em}
     \begin{subfigure}[b]{0.55\textwidth}
         \centering
         \includegraphics[width=\textwidth]{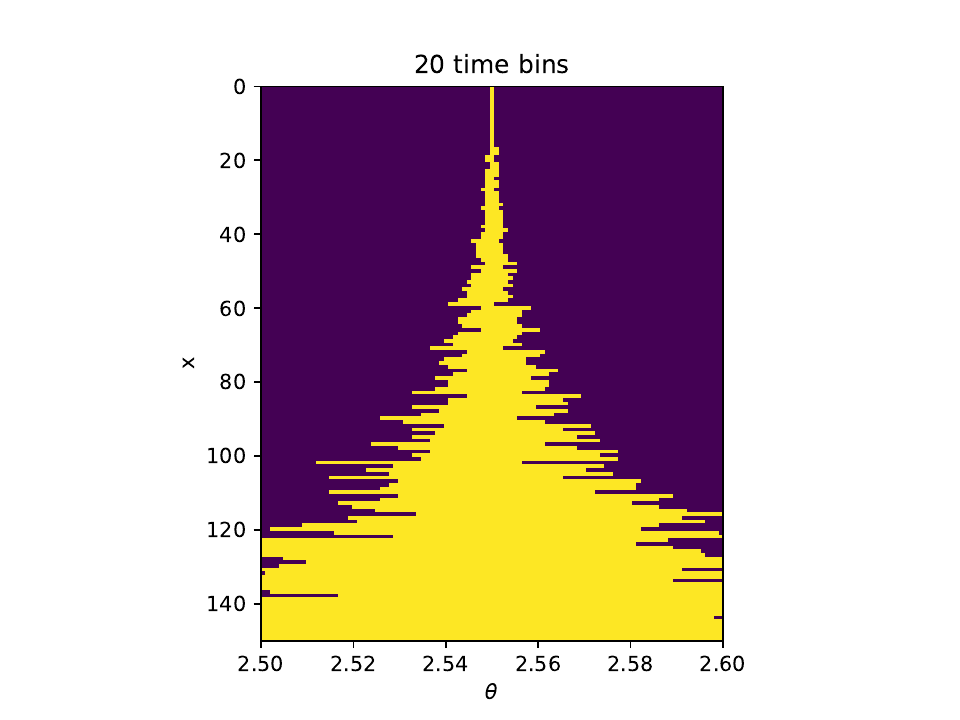}
         \label{subfig:20t_loglike_surface}
     \end{subfigure}
     \begin{subfigure}[b]{0.55\textwidth}
         \centering
         \includegraphics[width=\textwidth]{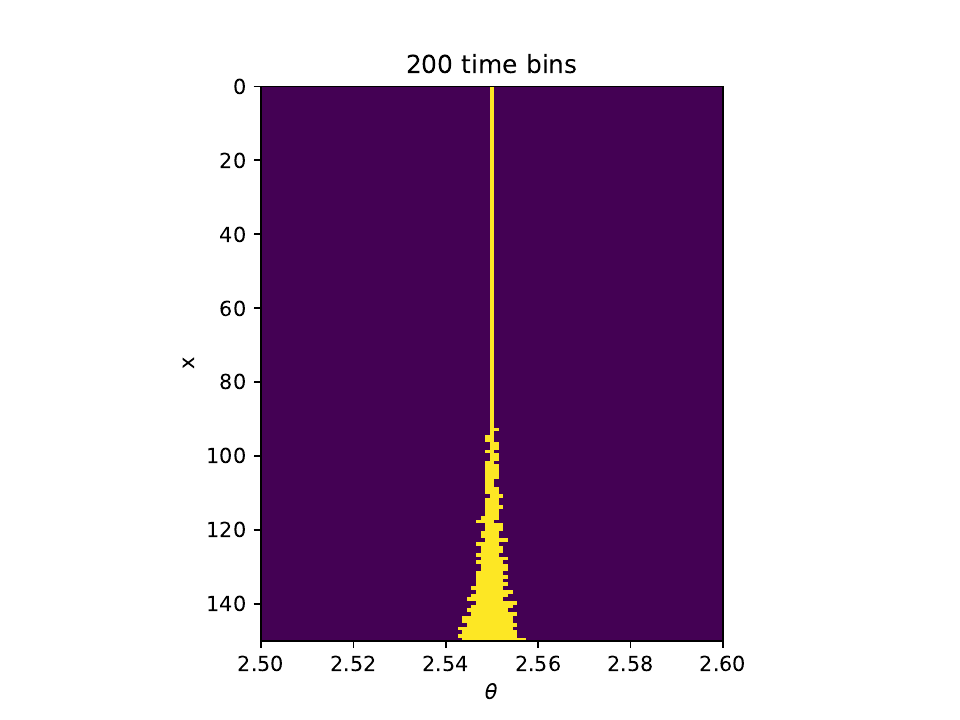}
        \label{subfig:200t_loglike_surface}
     \end{subfigure}
     \hspace{-7em}
     \begin{subfigure}[b]{0.55\textwidth}
         \centering
         \includegraphics[width=\textwidth]{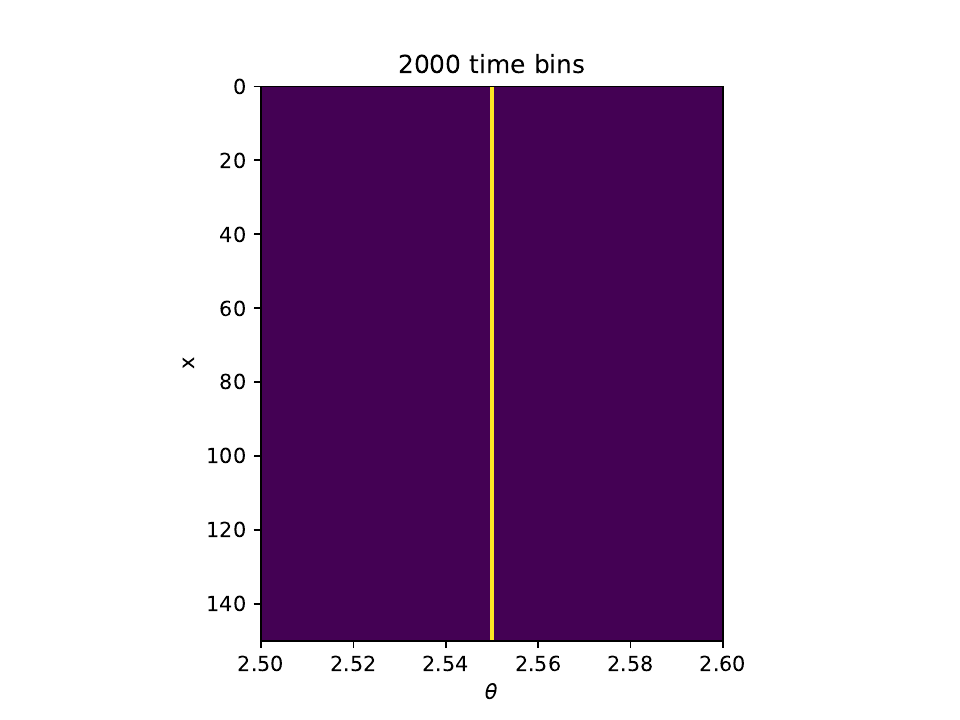}
         \label{subfig:2000t_loglike_surface}
     \end{subfigure}
        \caption{Plots of the time averaged likelihoods $\mathcal{L}_{ij}$ for simulated data sets generated according to \Cref{eq:data_toy_model} with 2, 20, 200 and 2000 time bins. The parameter and $x$ values that give likelihoods above a fixed threshold of $p=1e-3$ are shown in yellow and those that give likelihoods below the threshold are shown in purple, demonstrating that larger regions of parameter space fall below a fixed likelihood threshold as the number of time bins increases.}
        \label{fig:flagging_plots}
\end{figure*}

\Cref{fig:flagging_plots} shows the regions where the value of $\mathcal{L}_{ij}$ averaged over time falls above or below a threshold of $p=1e-3$, which is a typical value as determined from \citet{leeney22}. It can be seen that as the number of time bins increases, increasingly large portions of the space fall below the threshold, even in absence of anomalies, and so are flattened. By 2000 time bins, only a very small region of the parameter space is not flattened, which makes performing a model fit almost impossible.

This demonstrates that, especially in the case of time separated data, having an externally fixed threshold value can impede model fitting. Therefore, ideally the value of $p$ should be dynamic, allowing it to be low in suboptimal regions of the parameter space to avoid overflagging and flattening the likelihood surface, and higher around optimal regions to avoid missing genuine anomalous points. 

Therefore, as was suggested in \citet{leeney22}, this issue can be resolved by fitting the value of $p$ as a free parameter, rather than assigning it a fixed value and giving it a wide prior down to effectively zero. This results in there being sections of the parameter volume where the threshold is low and so the variations in the likelihood surface with the other parameters are visible. As a result, the model fit can progress, optimising the parameter values towards their posteriors while simultaneously optimising the threshold towards the optimal posterior value of the probability that a point is anomalous for that data set. This resolves the issue described above and allows the fit to proceed while still accurately flagging anomalous points. In the next section, tests of the entire process will be performed to demonstrate it's efficacy. Throughout the rest of this work, $p$ is fit as a free parameter with a logarithmically uniform prior in the range[1e-30 - 1e-3]. The next section will demonstrate the functionality of this method.

\subsection{Results}\label{sec:RFI_flag_results}
\begin{figure}
    \centering
    \includegraphics[width=\columnwidth]{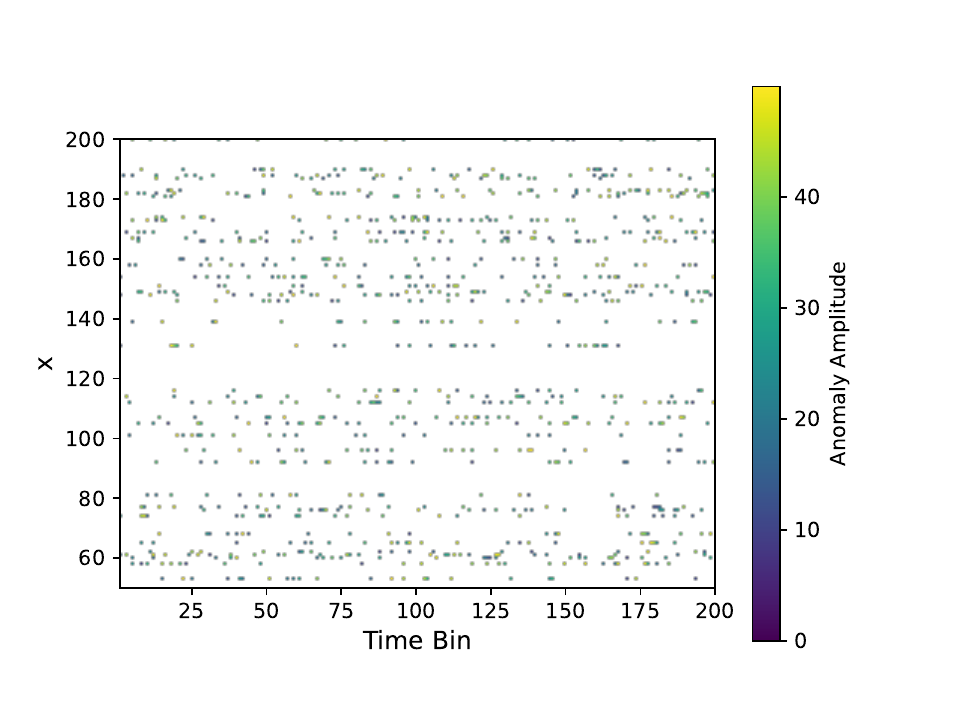}
    \caption{Plot of the anomalous points added to the $N_\mathrm{t}=200$ uncontaminated data set in order to produce the corresponding contaminated data set. The amplitudes were uniformly randomly chosen from the range 10-50. The time bin locations were uniformly randomly chosen and the $x$ bin locations were uniformly set to one of 40 randomly chosen channels.}
    \label{fig:RFI_example}
\end{figure}

In order to demonstrate the performance of this anomaly mitigation procedure, four test data sets were generated according to the model described in \Cref{sec:toy_model}, $N_\mathrm{t}$ = 2, 20, 200 and 2000 time bins, respectively. These data sets will henceforth be referred to as the \textit{uncontaminated} data. To each of these data sets, a random selection of anomalous peaks were added. In each case, the number of spikes added was equal to $N_\mathrm{t}\times 5$, such that every data set is contaminated in equal proportion. The amplitudes were uniformly randomly chosen from the range 10-50. The time bins in which each anomaly was placed were uniformly randomly chosen and the $x$ bin was uniformly set to one of 40 randomly chosen channels. This constrained the contamination to a subset of the data channels, in a manner more resembling RFI. This will be discussed in more detail in \Cref{sec:overcontam}. \Cref{fig:RFI_example} shows an example of the anomalous points added to the data for $N_\mathrm{t}=200$. These data sets will henceforth be referred to as \textit{contaminated}.

\subsubsection{Anomaly Correction}\label{sec:performance}
Two models were fit to each of the data sets described in the previous section. The first was a direct fit of the model given in \Cref{eq:toy_model_G} and \Cref{eq:toy_model_F}, with no attempt to correct for any anomalies. The second was a full fit of this model together with the time-dependent Bayesian anomaly flagging method described in the previous section, including the likelihood reweighting process and fitting for the threshold value as a parameter.

\begin{figure*}
    \centering
    \includegraphics[width=\textwidth]{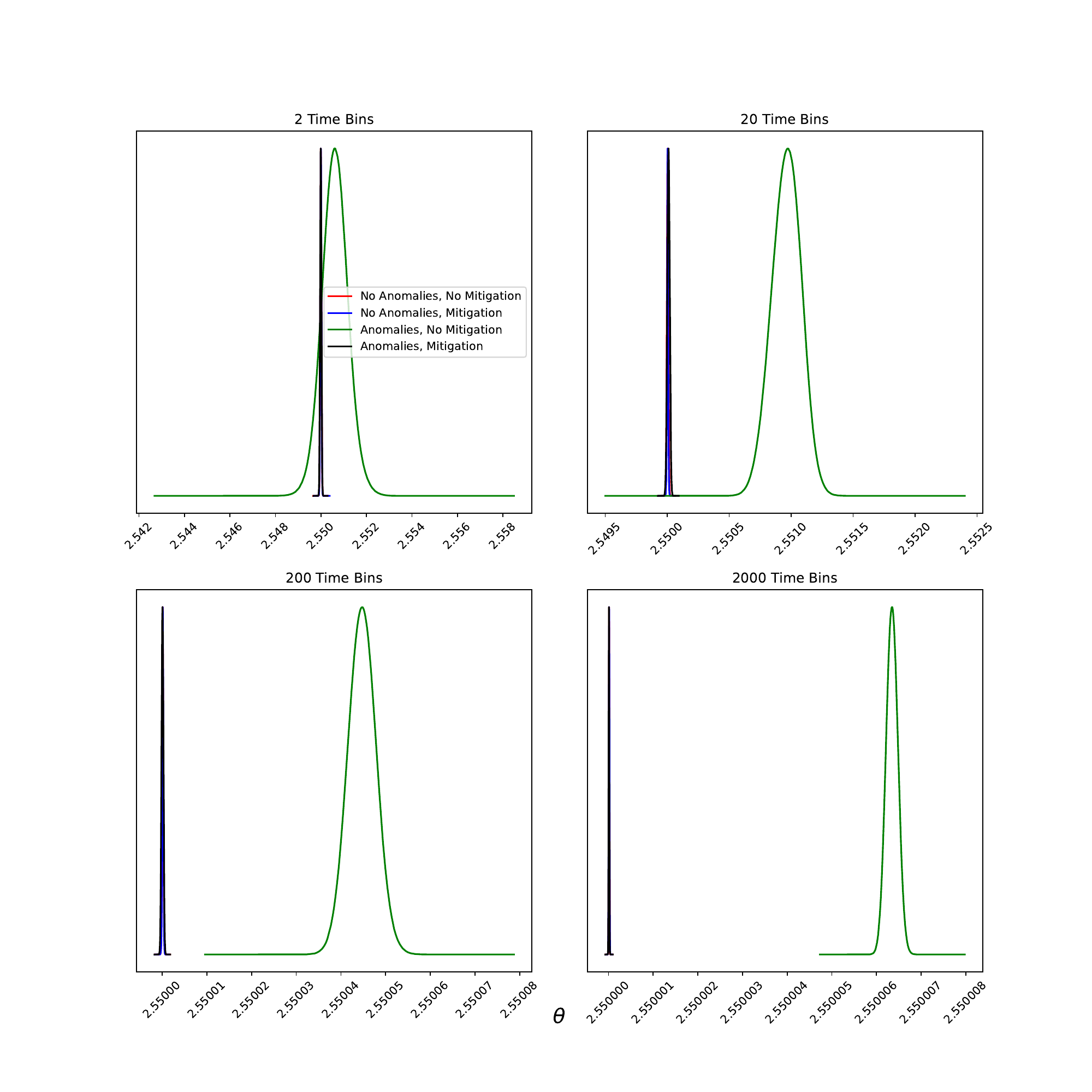}
        \caption{Plots of the posteriors on the toy model parameter $\theta$, when fitting both uncontaminated and contaminated test data sets with models that include  and don't include time-dependent Bayesian anomaly correction. The test data sets used a true value of $\theta =2.55$. Each subfigure shows the results for a simulated data set with a different number of time bins, from 2 to 2000. This demonstrates that uncorrected anomalies lead to parameter biases which are corrected by the application of the anomaly mitigation process.}
        \label{fig:channelled_anomaly_posteriors}
\end{figure*}

\Cref{fig:channelled_anomaly_posteriors} shows the posteriors on the parameter $\theta$ for each of these fits. For each number of time bins, the results of fitting the uncontaminated data sets without including any anomaly corrections, are shown in red in order to provide a benchmark. In all cases the posterior correctly identifies the true parameter value of 2.55, with the posterior becoming more tightly constrained around that value for greater numbers of time bins.

The posteriors when the correction is applied but the data is uncontaminated are shown in blue. In these cases, it can be seen that the recovered posteriors are highly consistent with that of the uncorrected cases. This demonstrates that including the time-dependent Bayesian anomaly correction method does not bias the fit in absence of any anomalies. This is the expected result.

The posteriors generated from contaminated data sets, but with no correction applied are shown in green. It can be seen that in all cases, the parameter posterior is biased from the true value. For $N_\mathrm{t}=2$, 20, 200 and 2000, the true value falls at $\sigma=1.1$, $\sigma=8.6$, $ \sigma=15.7$ and $\sigma=51.2$ respectively. This is again the expected result, demonstrating that the results will be biased if contamination is present but not accounted for. The offset increases with the number of time bins due to the high time bin cases having narrower posteriors, which makes the bias more apparent.  

The posteriors when the correction is applied to contaminated data are shown in black. It can be seen that the proposed time-dependant Bayesian anomaly mitigation methodology has successfully countered the bias in the posterior seen in the contaminated but uncorrected cases. For $N_\mathrm{t}=2$, 20, 200 and 2000, the true value now falls at $\sigma=0.42$, $\sigma=1.50$, $\sigma=0.34$ and $\sigma=1.60$ respectively. The true value of the parameter is therefore now recovered to the $2\sigma$ level in all cases.

This demonstrates that the proposed anomaly mitigation technique is successfully correcting for the added anomalies in the data and enabling the model parameters to be recovered accurately. It is worth noting that in this test case the simulated data is heavily contaminated, with 3.3\% of data points featuring an anomaly and 26.5\% of $x$ channels being contaminated to some extent. Despite this, anomalies are still correctly accounted for.

\subsubsection{Anomaly Recovery}\label{sec:anom_rec}
\Cref{fig:channelled_anomaly_posteriors} demonstrates that our methods account for the presence of anomalous data points to a level sufficient that the underlying model can be accurately recovered. It is also worth assessing directly whether the points predicted to be anomalies correctly correspond to those added into the data.

This was investigated for the four corrected fits to contaminated data shown in black in \Cref{fig:channelled_anomaly_posteriors}. For each case, the full slow likelihood was applied to each sample of the fit's final posterior, in order to determine which data points were being flagged as anomalous for each posterior sample and the weighted average was taken. This determines the points that were predicted as anomalies in the final likelihood. Then the amplitude of these anomalous points was calculated by taking the residual of the input data to the weighted posterior average model at those points. 

\begin{figure*}
    \centering
    \includegraphics[width=\textwidth]{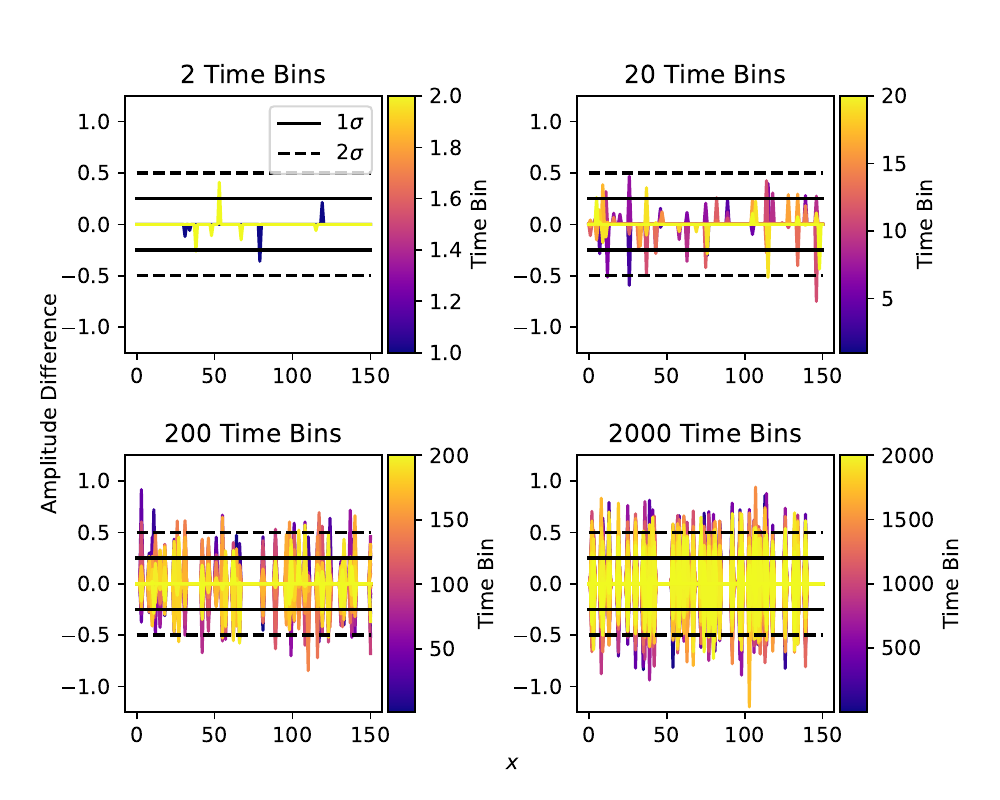}
    \caption{Plot of the differences between the anomalous points input to the simulated data sets and those identified by the anomaly correction process for the four fits shown in black in \Cref{fig:channelled_anomaly_posteriors}. The 1 and 2 $\sigma$ noise levels are marked to demonstrate that the identified anomalous points differ from the true anomalies only by expected Gaussian noise.}
    \label{fig:recovered_anomaly_comp}
\end{figure*}

\Cref{fig:recovered_anomaly_comp} shows the difference between the anomalies the algorithm identified to those that were originally added to the data. It can be seen that for all four cases, all anomalies added into the data, which had a minimum amplitude of 10, are successfully identified by this process, and their amplitude is correctly determined to within the expected noise error of a normal distribution with $\sigma=0.25$, which was the level of noise added to the toy data. 
This accuracy of anomaly recovery, even for these heavily contaminated toy models, raises the possibility that our anomaly correction methodology could also be used to detect anomalous points of interest, and thus function as a transient flagger. This possibility will be explored further in \Cref{sec:transient_detection}.

\subsubsection{Computation Time}\label{sec:comp_time}
Th motivation for implementing likelihood reweighting as described in \Cref{sec:LRW} was to improve the computational efficiency of the proposed process and significantly reduce the otherwise strong dependency of the total runtime on the number of time bins.

In order to investigate the effects on runtime, the four contaminated data sets, with $N_\mathrm{t} = 2$, 20, 200 and 2000, were fit to the corresponding model with the anomaly correction method implemented, but without using likelihood reweighting. Instead the full `slow' likelihood defined in \Cref{eq:time_sep_flagged} was used. 

\begin{figure}
    \centering
    \includegraphics[width=\columnwidth]{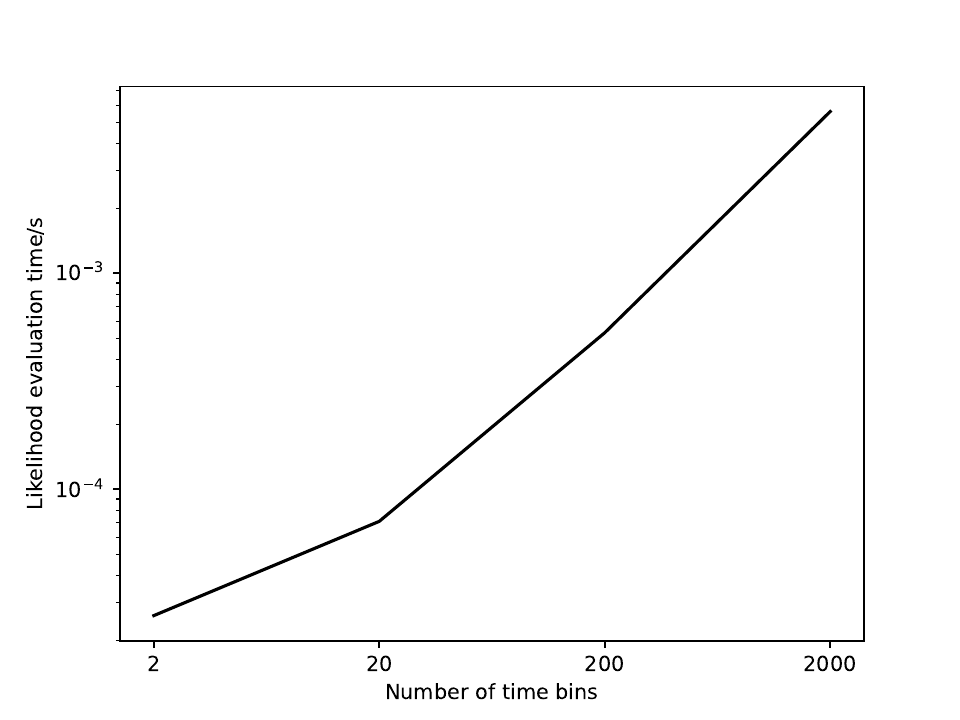}
    \caption{Plot of the time required for a single likelihood evaluation of the full time-dependent anomaly mitigation method described in \Cref{eq:time_sep_flagged}.}
    \label{fig:slow_L_runtimes}
\end{figure}

\Cref{fig:slow_L_runtimes} shows the time for a single likelihood evaluation for each of these fits. It can be seen that the evaluation time increases proportionally with the increasing number of time bins, as expected.

\begin{figure}
    \centering
    \includegraphics[width=\columnwidth]{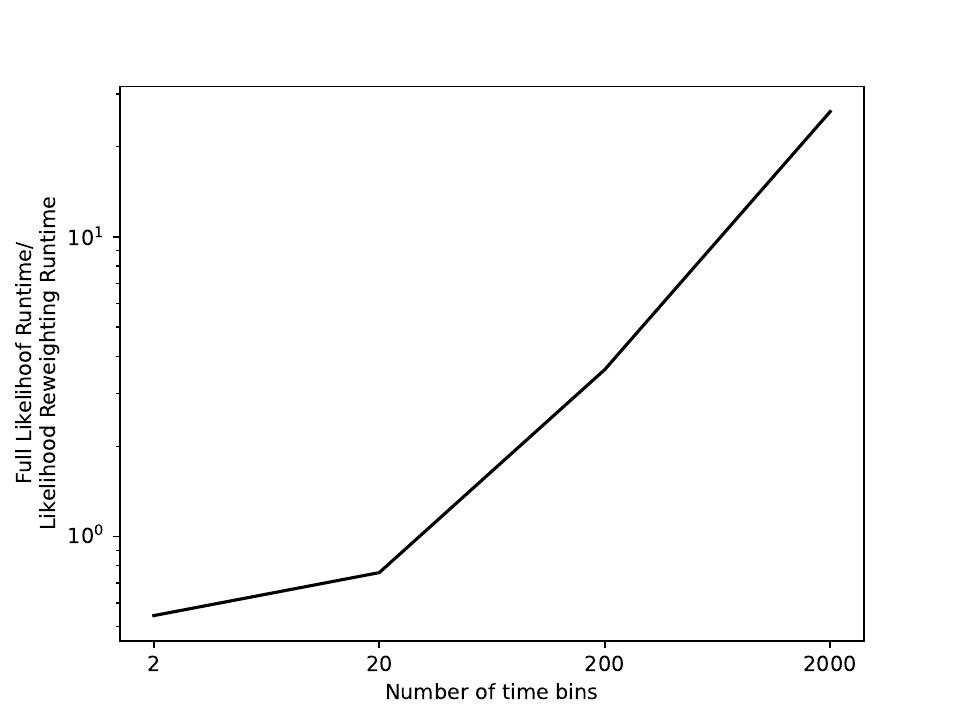}
    \caption{Plot of the ratio of the total runtime of a model fit implementing time-dependent Bayesian anomaly flagging using the full slow likelihood described in \Cref{eq:time_sep_flagged}, to that of the equivalent fit with using likelihood reweighting.}
    \label{fig:runtime_comp}
\end{figure}

\Cref{fig:runtime_comp} shows the ratio of the total runtime of the fit with no likelihood reweighting to an equivalent fit on the same data set with likelihood reweighting implemented. It can be seen that implementing likelihood reweighting reduces the runtime when $N_\mathrm{t}>20$, with the speed up following approximately the same trend as that of likelihood evaluation seen in \Cref{fig:slow_L_runtimes}. For example, for 2000 time bins, the run time improves by a factor of 25. This demonstrates that implementing likelihood reweighting successfully makes the total runtime close to independent of the number of time bins and thus enables the flagging process to be implemented efficiently on very large data sets, as was the objective. 

\subsubsection{Overcontamination}\label{sec:overcontam}
In \Cref{sec:comp_time}, it was demonstrated that the process of likelihood reweighting described in \Cref{sec:LRW} is successful at significantly improving the computation time for the proposed method. However, it also introduces a limitation that is not present if the full, slow likelihood is used. 

When likelihood reweighting is implemented for this process, the initial model fit is performed using the `fast' likelihood defined in \Cref{eq:fast_L}. However, in this fast likelihood, in order for the evaluation time to not be dependent on the number of time bins as required, anomalies are identified by comparing the time-average of the model to averaged data and flagging the entire data channel if the likelihood is below the specified threshold. However, as a result of this, if the data is contaminated in such as way that most or all $x$ channels are contaminated to some extent, the baseline of the average data will be offset, resulting in a failure to flag this low level `background' contamination, which is instead fit with the parameterised model. This can therefore result in a bias in the fit model.

\begin{figure}
    \centering
    \includegraphics[width=\columnwidth]{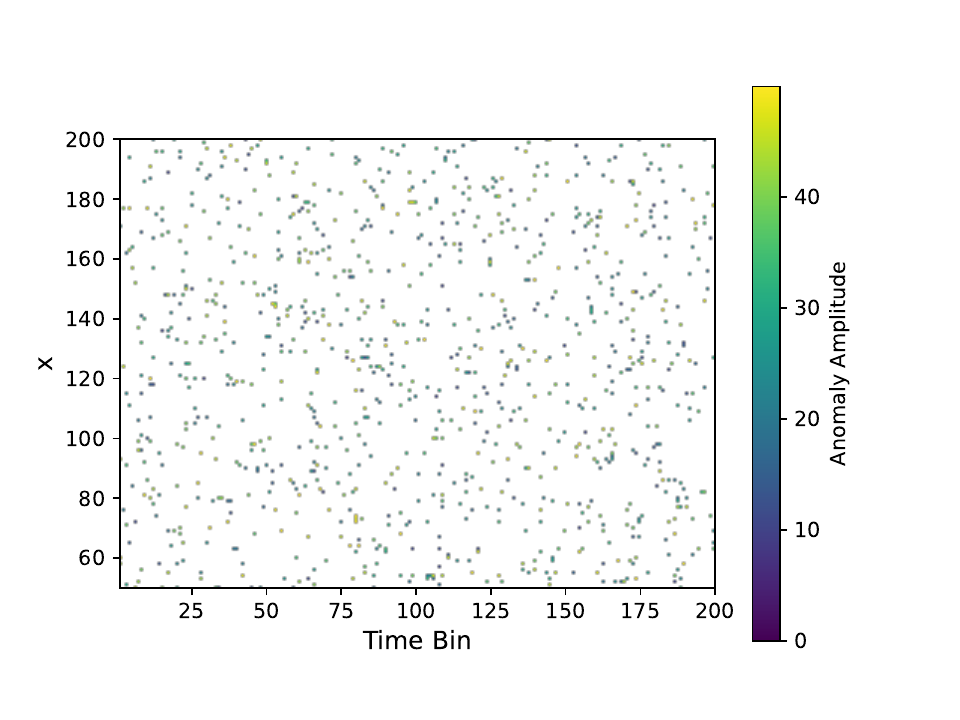}
    \caption{Plot of the anomalous points added to the $N_\mathrm{t}=200$ uncontaminated data set in order to produce the corresponding contaminated data set in which most $x$ channels are contaminated to some degree. The amplitudes were uniformly randomly chosen from the range 10-50. The time bin and $x$ bin locations were uniformly randomly chosen.}
    \label{fig:random_RFI_example}
\end{figure}

In order to test this effect, a new set of contaminated data sets were generated, by adding anomalous data points to the four uncontaminated data sets described in \Cref{sec:RFI_flag_results}, in exactly the same manner as previously, except the $x$ bin of each contaminated point was chosen entirely randomly, rather than being confined to certain channels. \Cref{fig:random_RFI_example} shows an example of the anomalous points added to the data for $N_\mathrm{t} = 200$.

\begin{figure*}
    \centering
    \includegraphics[width=\textwidth]{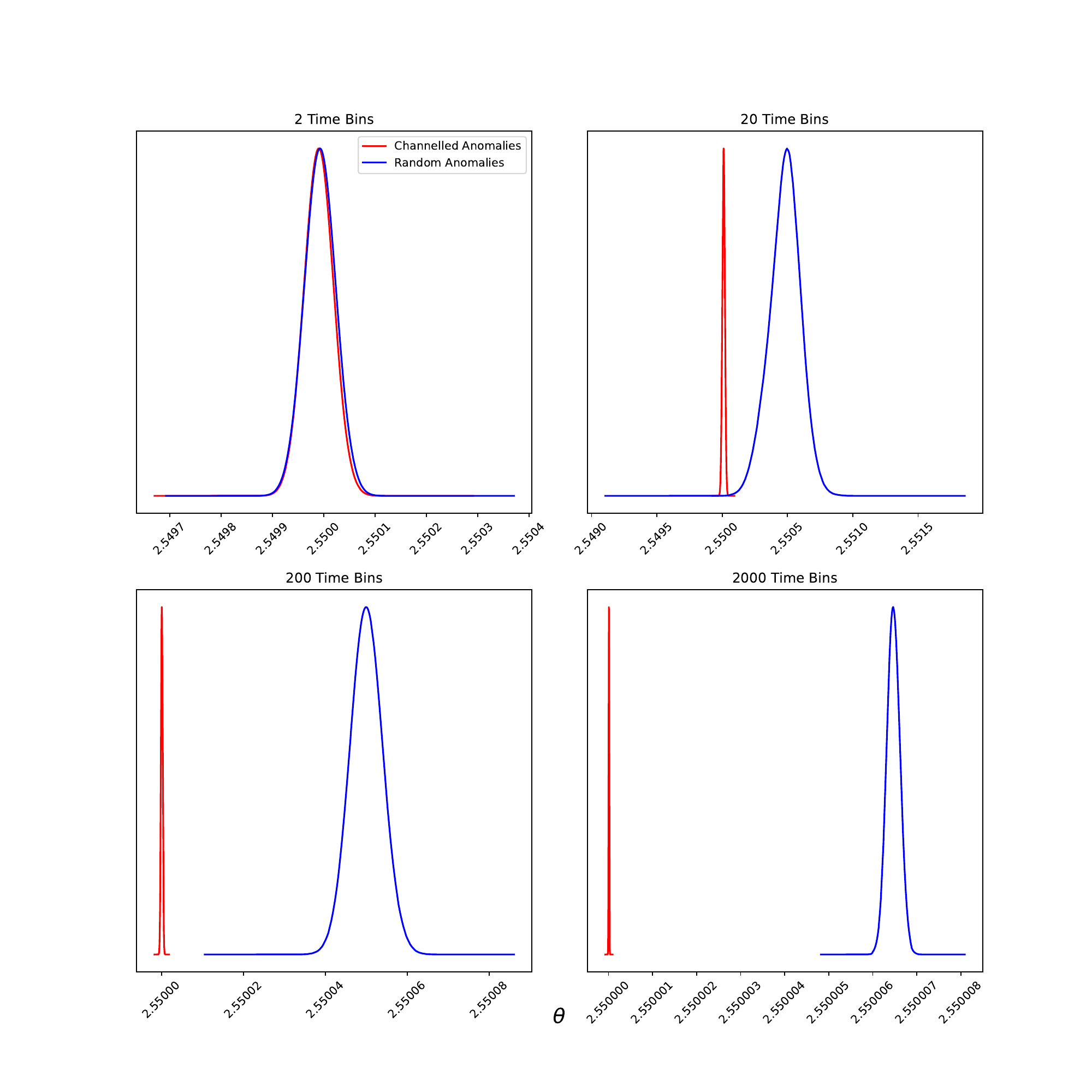}
    \caption{Plot of the parameter posteriors recovered when applying the Bayesian anomaly mitigation method with likelihood reweighting to simulated test data sets contaminated with $5N_t$ anomalous points located at random (blue), in comparison to the recovered posteriors for performing the same fits on data instead contaminated with the same number of anomalous points but constrained to a maximum of 40 $x$ channels (red), previously shown in \Cref{fig:channelled_anomaly_posteriors} (in black).}
    \label{fig:channelled_random_posterior_comp}
\end{figure*}

The tests described in \Cref{sec:performance} were repeated on these new randomly contaminated data sets. \Cref{fig:channelled_random_posterior_comp} shows the posteriors for the cases where the contaminated data sets were fit with the anomaly correcting likelihood, in comparison to the equivalent fits where the anomalies were confined to specific channels.

Given that the number of anomalous points injected was $5 \times N_\mathrm{t}$ and the number of $x$ channels in the simulated data sets are 151, for fully randomly distributed anomalies, the expected number of anomalous points per channel will be 0.07, 0.66, 6.62 and 66.23 for $N_\mathrm{t}=$ 2, 20, 200 and 2000 respectively. For all cases except $N_\mathrm{t}=2$, it is more likely that a channel be contaminated than not, with $N_\mathrm{t}=200$ and $2000$ expecting multiple contaminated points per channel. It can therefore be expected that these model fits would exhibit a bias in the recovered posterior. It can be seen in \Cref{fig:channelled_random_posterior_comp} that, as the number of time bins increases, the parameter recovery when the anomalies are random becomes increasingly biased to increasing sigma levels of $\sigma=0.3$, $\sigma=4.3$, $\sigma=14.7$ and $\sigma=58.3$. As expected, the $N_\mathrm{t}=2$ case, which does not exhibit overcontamination, does not show a bias. For the remaining cases, it can be seen that the offset of the posterior mean actually decreases as $N_\mathrm{t}$ increases, but the posterior width decreases at a faster rate, resulting in the observed increasing bias.

This overcontamination bias can be overcome by utilising the full, slow likelihood instead of implementing likelihood reweighting. However, doing so would prevent the improved calculation time the reweighting provides. This, therefore represents a limitation of the proposed methodology. However, the excessive random contamination required for this limitation to become relevant is quite physically unreasonable, particularly in the context of RFI as it requires approximately more than half the channels to show contamination. Therefore, this limitation is of minor concern.

\section{Transient Detection}\label{sec:transient_detection}
In \Cref{sec:anom_rec}, it was demonstrated that as well as enabling accurate identification of the underlying model from beneath contamination, our methods also give accurate recovery of the anomalous points themselves. This raises the possibility that, in addition to applications in RFI excision, our methods could also be applied to detect transient signals of interest. As the implementation of likelihood reweighting enables model fitting to have a computational time almost independent of the number of time bins, this process potentially provides a Bayesian methodology for efficiently searching large data sets for transients such as fast radio bursts and pulsars.  

\subsection{Anomaly Recovery}\label{sec:random_recovery}
In order to test the ability of our methods to correctly identify small transients in large data sets, a series of tests were run in which a single anomalous data point was added at random to the previously described 2000 time bin test data set, and the model fit with the flagging process implemented. This was repeated 5 times each for anomalous points with amplitudes of 0.5, 1.25, 2.5, 5 and 12.5, which correspond to SNRs of 2, 5, 10, 20 and 50 respectively.

\begin{figure}
    \centering
    \includegraphics[width=\columnwidth]{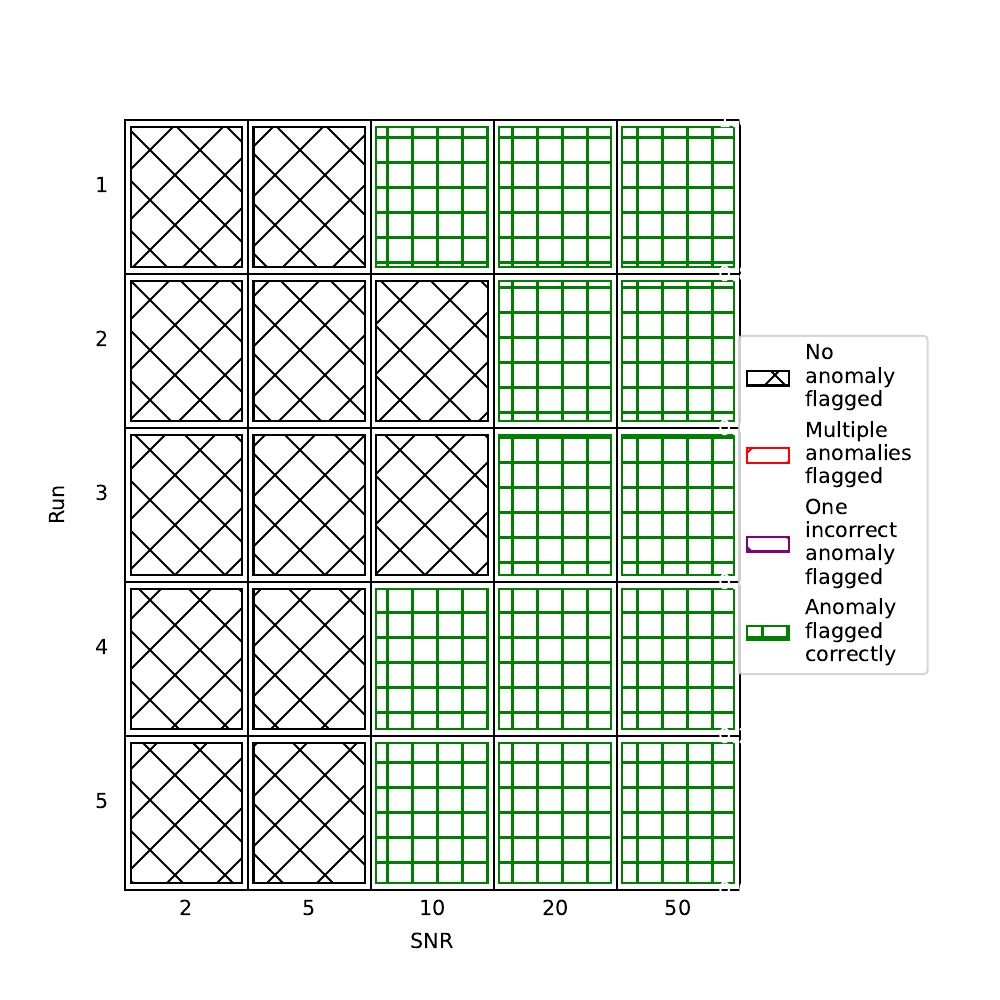}
    \caption{Plot of the attempts to identify a single added anomalous point into 2000 time bin data sets using the proposed method as a transient detector. Each panel shows the results for different SNR of the added anomalous point and for different repeats with the anomaly in a different random data bin. Cases where no anomalies were identified at all are shown in black. Cases where the single anomaly was correctly flagged are shown in green. Cases where multiple points were erroneously flagged and cases where only one point was flagged, but at the wrong location would be shown in red and purple respectively. However, no such cases were seen.}
    \label{fig:transient_results}
\end{figure}

The result of these fits are shown in \Cref{fig:transient_results}. It can be seen that when acting as a flagger, our methods correctly identify the single anomalous point for the cases with higher SNR. Furthermore, they never erroneously identifies any data points as anomalous that were not anomalous in the data. However, the flagger does begin to fail to detect the anomalous point at lower SNRs, which can be expected given there is only one anomalous point out of 302,000 in this test case, meaning lower SNR anomalies can become indistinguishable from simple statistical fluctuations and so will not be flagged. 

This demonstrates the potential for the proposed methodology to function as an efficient Bayesian transient detector. This will be explored in greater detail in a future work.

\section{Conclusions}\label{sec:conclusions}
RFI is a significant challenge in radio astronomy. In this paper, we extend the Bayesian RFI mitigation methodology presented in \citet{leeney22} into the time domain. This enables transient anomalies to be flagged and properly accounted for in a Bayesian manner when fitting models to time-series data. 

The process of likelihood reweighting was implemented in order to enable this process to be performed in a manner mostly independent of the number of time bins in the data. This was demonstrated to produce significant improvements in the computation time as the number of time bins increases, reaching a 25 times speed increase on a test case with 2000 time bins, by breaking the proportional relation between the number of time bins and the runtime.

Our methodology was demonstrated to be successful at correcting for contamination in a series of test data sets that approximate global 21cm experiment data. It accurately corrected the bias in the model parameters which occurs if the contamination was not accounted for, whilst not affecting the results if no contamination is present. 

Furthermore, it was demonstrated that our methods can correctly extract anomalous points from data. Therefore, the efficacy of our methods as an efficient transient detector was explored. It was demonstrated that they were successfully able to identify a single anomalous point out of 302,000, provided the anomaly had a SNR of 10 or higher. The use of this process as a transient flagger will be explored in greater depth in a future work.

A potential limitation was also identified, in which the implementation of likelihood reweighting results in a failure to correctly account for anomalies in cases where every data channel is contaminated to some degree. However, this situation is unlikely to be encountered in practice.

Overall, the methodology presented here represents an efficient and fully Bayesian technique for correcting for time-dependent contamination or identifying transients in large data sets.

\section*{Acknowledgements}
We would like to thank Will Handley for his work on the development of the original methodology.
Dominic Anstey and Sam Leeney were supported by the Science and Technologies Facilities Council.

\section*{Data Availability}
The data used and generated in this article will be shared on reasonable request to the corresponding author.

%%%%%%%%%%%%%%%%%%%%%%%%%%%%%%%%%%%%%%%%%%%%%%%%%%



\bibliographystyle{mnras}
\bibliography{bib} % if your bibtex file is called example.bib

%\begin{thebibliography}{99}
%\end{thebibliography}

%%%%%%%%%%%%%%%%%%%%%%%%%%%%%%%%%%%%%%%%%%%%%%%%%%

%%%%%%%%%%%%%%%%% APPENDICES %%%%%%%%%%%%%%%%%%%%%

%\appendix



%%%%%%%%%%%%%%%%%%%%%%%%%%%%%%%%%%%%%%%%%%%%%%%%%%

% Don't change these lines
\bsp	% typesetting comment
\label{lastpage}
\end{document}